\documentclass[pdflatex,sn-mathphys-num]{sn-jnl}
\usepackage{graphicx}%
\usepackage{multirow}%
\usepackage{amsmath,amssymb,amsfonts}%
\usepackage{amsthm}%
\usepackage{mathrsfs}%
\usepackage[title]{appendix}%
\usepackage{xcolor}%
\usepackage{textcomp}%
\usepackage{manyfoot}%
\usepackage{booktabs}%
\usepackage{algorithm}%
\usepackage{algorithmicx}%
\usepackage{algpseudocode}%
\usepackage{listings}%
\theoremstyle{thmstyleone}%
\theoremstyle{thmstyletwo}%

\theoremstyle{thmstylethree}%

\usepackage{amsmath}
\usepackage{amssymb}
\usepackage{physics}
\usepackage{graphicx} 
\usepackage{enumerate}
\usepackage[inline]{enumitem}
\usepackage{bm}
\usepackage{tcolorbox}
\usepackage{tikz}
\usepackage{listings}

\usepackage{hyperref}
\hypersetup{hidelinks,
  colorlinks=true,
  allcolors=blue,
  pdfstartview=Fit,
  breaklinks=true
}

\begin{document}

\title[Article Title]{Scalable Neural Quantum State based Kernel Polynomial Method for Optical Properties from the First Principle}

\author[1]{Wei Liu}

\author[1]{Rui-Hao Bi}

\author*[1,2,3]{Wenjie Dou}\email{douwenjie@westlake.edu.cn}

\affil*[1]{Department of Chemistry, School of Science and Research Center for Industries of the Future, Westlake University, Hangzhou, Zhejiang 310030, China}

\affil[2]{Institute of Natural Sciences, Westlake Institute for Advanced Study, Hangzhou, Zhejiang 310024, China}

\affil[3]{Key Laboratory for Quantum Materials of Zhejiang Province, Department of Physics, School of Science and Research Center for Industries of the Future, Westlake University, Hangzhou, Zhejiang 310030, China}


\abstract{
Variational optimization of neural-network quantum state representations has achieved FCI-level accuracy for ground state calculations, yet computing optical properties involving excited states remains challenging. In this work, we present a neural-network-based variational quantum Monte Carlo approach for ab-initio absorption spectra. We leverage parallel batch autoregressive sampling and GPU-supported local energy parallelism to efficiently compute ground states of complex systems. By integrating neural quantum ground states with the kernel polynomial method, our approach accurately calculates absorption spectra for large molecules with over 50 electrons, achieving FCI-level precision. The proposed algorithm demonstrates superior scalability and reduced runtime compared to FCI, marking a significant step forward in optical property calculations for large-scale quantum systems.
}

\maketitle

The central challenge of ab-initio quantum chemistry~\cite{friesner2005ab,hermann2023ab} lies in accurately modeling optical properties of complex systems. These properties, such as absorption and emission spectra, depend on precise descriptions of excited states, which are significantly harder to compute than ground states. With the advent of ultra-fast pump-probe spectroscopy~\cite{bv2000absorption}, experimental probing of excited states has become possible, increasing the demand for efficient and accurate electronic structure methods capable of handling systems with large numbers of electrons.

The absorption spectrum can be directly calculated using excited state information as~\cite{zuehlsdorff2019modeling}:
\begin{equation}
    I(\omega) = \sum_n |\expval{n|\hat{\mu}|0}|^2\delta(\omega_n-\omega),
    \label{eqn:Iw2}
\end{equation}
where $\ket{0}$ is the ground state, $\omega_n$ and $\ket{n}$ are the $n$-th excited state energy and eigenvector, and $\hat{\mu}$ is the dipole moment operator. The commonly used Hamiltonian in quantum chemistry is the ab-initio electronic Hamiltonian under the Born-Oppenheimer approximation~\cite{born1927quantentheorie}, given by: 
\begin{equation}
    \hat{H}_{\text{el}}=-\frac{1}{2}\sum_i \nabla_i^2-\sum_{i,I}\frac{Z_{iI}}{|\mathbf{r}_i-\mathbf{R}_I|}+\frac{1}{2}\sum_{i\neq j}\frac{1}{|\mathbf{r}_i-\mathbf{r}_j|}.
    \label{eqn:boh}
\end{equation}
This includes the kinetic energy of electrons, Coulomb attraction between electrons and nuclei, and Coulomb repulsion between electrons. Practical representations of $\hat{H}_{\text{el}}$ use finite basis sets, introducing discretization errors, though stochastic methods like quantum Monte Carlo (QMC)~\cite{foulkes2001quantum} can reach the complete basis set limit. The task is to compute the ground and excited state energies and eigenvectors of $\hat{H}_{\text{el}}$. For a given basis set, the time-independent Schrödinger equation (TISE)~\cite{szabo1996modern} $\hat{H}\ket{\Psi}=E\ket{\Psi}$
  exactly for $\hat{H}_{\text{el}}$ can be solved exactly in the many-electron Hilbert space using full configuration interaction (FCI)~\cite{knowles1984new} or exact diagonalization (ED)~\cite{weisse2008exact}. FCI provides exact absorption spectra via Eq.~\ref{eqn:Iw2}, but its exponential scaling limits applications to systems with 15 to 16 electrons, too small for most chemical systems. For larger systems, approximations like multi-reference configuration interaction (MRCI)~\cite{szalay2012multiconfiguration} or time-dependent density functional theory (TDDFT)~\cite{marques2004time} are used, sacrificing accuracy.

Within linear response theory~\cite{kubo1991statistical} and the Kubo formalism~\cite{kubo1957statistical,kubo1957statistical2}, molecular absorption spectra are expressed through the Fourier transform of dipole-dipole correlation functions~\cite{kubo1966fluctuation,liu2023predicting,liu2024memory}:
\begin{equation}
    I(\omega) \propto \int_{-\infty}^{\infty}C_{\hat{\mu}\hat{\mu}}(t) e^{i\omega t} dt,
\end{equation}
where $C_{\hat{\mu}\hat{\mu}}(t)=\expval{\hat{\mu}(t)\hat{\mu}(0)}$ is the dipole autocorrelation function and $\hat{\mu}(t)$ is the dipole operator in Heisenberg picture. The autocorrelation function can be evaluated with 
the time-dependent quantum state $\ket{\Psi(t)}$, governed by the time-dependent Schrödinger equation (TDSE)~\cite{iitaka1994solving},
\begin{equation}
    i\frac{\text{d}\ket{\Psi(t)}}{\text{d}t}=\hat{H}(t)\ket{\Psi(t)},
\end{equation}
with the time-dependent Hamiltonian $\hat{H}(t)$. While ab-initio variational wave functions have recently succeeded in real-time TDSE evolution for systems like a quantum dot with 18 electrons~\cite{nys2024ab}, limitations in accuracy and efficiency have restricted their broader application to many-electron systems, such as chemical molecules. By rescaling the Hamiltonian  ($\hat{H}\rightarrow \tilde{H}$) and energy ($\omega\rightarrow\tilde{\omega}$), the absorption spectrum can be efficiently computed using the kernel polynomial method (KPM)~\cite{weisse2006kernel},
\begin{equation}
    \text{Im}[C_{\hat{\mu}\hat{\mu}}(\tilde{\omega})]=-\frac{1}{\sqrt{1-\tilde{\omega}^2}}\bigg( \tilde{\Omega}_0 + 2\sum_{n=1}^{\infty} \tilde{\Omega}_nT_n(\tilde{\omega})\bigg),
    \label{eqn:kpm}
\end{equation}
where $T_n(\tilde{\omega})=\cos[n\arccos(\tilde{\omega})]$, $\tilde{\Omega}_n=g_n\Omega_n$, $g_n$ is the Jackson kernel (Eq.~\ref{eqn:jackson})~\cite{vyazovskaya2006normalizing}, and moments $\Omega_n$ are computed as:
\begin{equation}
    \Omega_n =\bra{0}\hat{\mu}T_n(\tilde{H})\hat{\mu}\ket{0}.
    \label{eqn:moments}
\end{equation}
Here, $\hat{H}$ is the time-independent Hamiltonian, and  $\ket{0}$ is its ground state of $\hat{H}$ (see Methods). Importantly, calculating the absorption spectrum only requires solving for the ground state $\ket{0}$, which satisfies TISE. Achieving FCI-level accuracy for $\ket{0}$ directly translates to FCI-level accuracy for absorption spectra, providing significant improvements in efficiency and storage.

As discussed, FCI scales exponentially with system size, prompting modern quantum chemistry to focus on approximations that balance accuracy and computational feasibility. The Hartree–Fock (HF) method~\cite{lykos1963discussion} treats electrons as independent particles influenced by the nuclei and a mean-field from other electrons, calculating expansion coefficients for atomic orbitals. Configuration interaction (CI) methods~\cite{shavitt1977method,sherrill1999configuration} use a restricted active space built from spin orbitals to approximate electron configurations. Coupled cluster (CC) approaches~\cite{bartlett2007coupled} employ an exponential cluster operator to model electron correlation, though they occasionally yield unphysical solutions. 
The density matrix renormalization group (DMRG) method~\cite{white1992density,white1993density} achieves FCI-level precision for molecular systems with up to 30 electrons~\cite{brabec2021massively,larsson2022chromium}, but its efficiency is limited by the matrix product state ansatz, which only captures finitely correlated states~\cite{perez2006matrix}. Additionally, DMRG is poorly suited for large-scale parallelization.  
Neural quantum states (NQS) based on restricted Boltzmann machines (RBM)~\cite{fischer2012introduction,choo2020fermionic} use neural networks to solve many-body quantum systems~\cite{carleo2017solving}. RBM has shown higher accuracy than CC with excitations up to second order CCSD and third order (CCSD(T)) for various molecular systems with up to 24 spin orbitals. Recently, autoregressive NQS architectures (NAQS)~\cite{barrett2022autoregressive,wu2023nnqs} have been introduced for ab-initio quantum chemistry, enabling exact sampling of physical states through neural autoregressive models~\cite{larochelle2011neural}. Subsequently, a parallelization scheme based on NAQS was implemented using a mask autoencoder based on distributed estimation (MADE)~\cite{zhao2023scalable}. These advancements have extended NQS to systems with more than $30$ electrons, surpassing RBM-based methods, but it is still limited in large-scale excited state applications.

To address these challenges, we propose a scalable NQS-KPM (sNQS-KPM) framework for ab-initio absorption spectra, enabling large-scale excited-state calculations. By combining the NQS with the KPM (Eqs.~\ref{eqn:kpm} and \ref{eqn:moments}), sNQS-KPM achieves efficient and accurate parallel computations. Our method significantly improves computational efficiency while maintaining accuracy comparable to ED. Using a parallel sampling strategy based on the autoregressive wavefunction ansatz, it ensures low sampling costs and balanced workloads. Additionally, we implement a batch-parallel local energy scheme with a compressed Hamiltonian representation, drastically reducing memory usage. Finally, we efficiently calculated the moments in KPM by using the recursive relationship, which avoided us from storing a large number of eigenvectors to calculate absorption spectra (Eq.~\ref{eqn:Iw2}), saving a lot of storage. These allow us to push the NQS towards the excited state era by calculating absorption spectra based on the neural quantum ground state. Experiments demonstrate its effectiveness on systems with up to 52 electrons and Hamiltonians containing millions of terms.

\begin{figure}[htbp]
    \centering
    \includegraphics[width=0.98\linewidth]{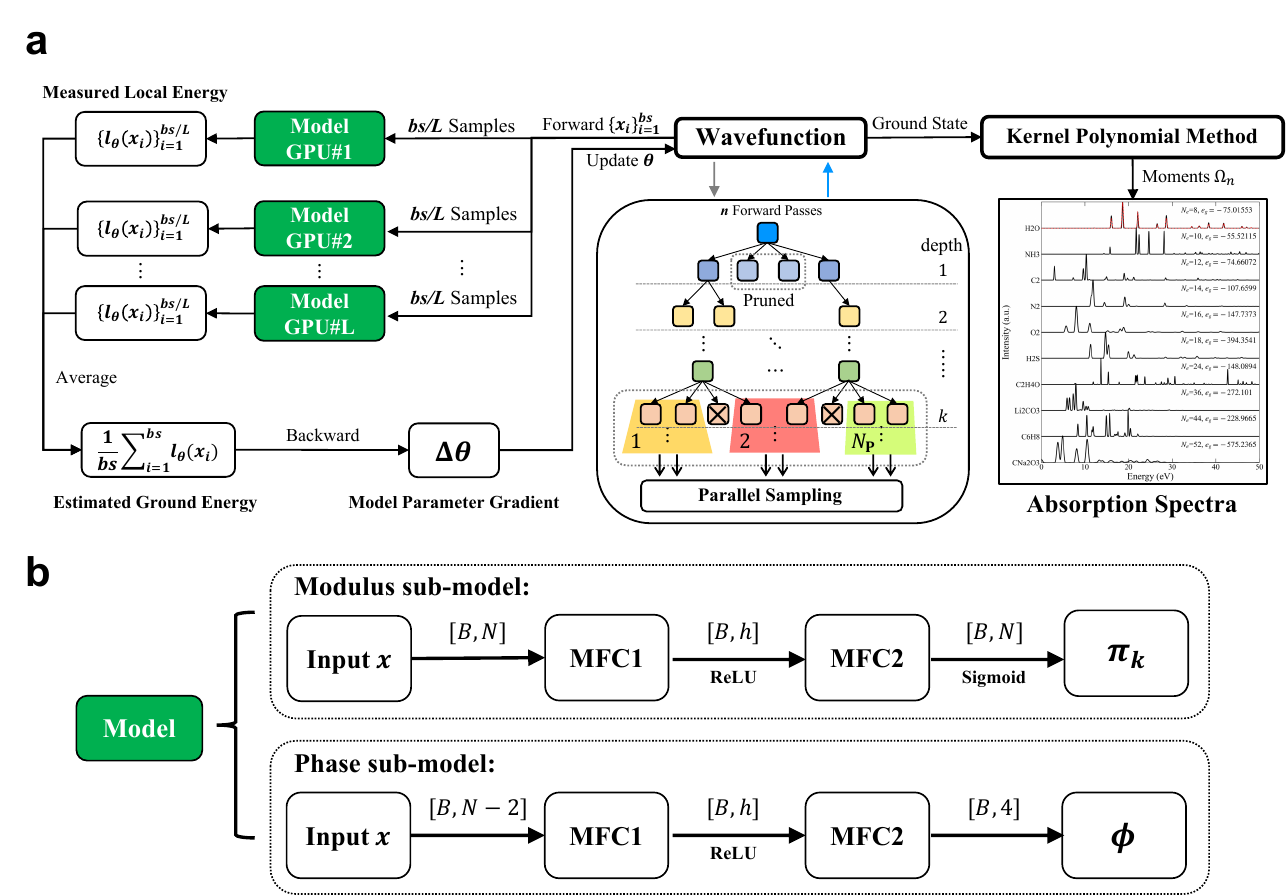}
    \caption{\textbf{Architecture of sNQS-KPM for ab-initio absorption spectra.} \textbf{a,} Overview of the sNQS-KPM algorithm, showing parallel VQMC optimization (left), parallel autoregressive sampling (middle), and KPM (right). \textbf{b,} Neural quantum state model architecture, splitting the wavefunction output into modulus and phase components, with batch size $B$, dimensions $N$, hidden layer size $h$, and masked fully connected (MFC) layers. Activation functions include ReLU and Sigmoid.}
    \label{fig:summary}
\end{figure}

\section{Results}

\textbf{Neural-network wavefunction ansatz.}
The architecture of sNQS-KPM for ab-initio absorption spectra is shown in Fig.~\ref{fig:summary}. 
 We decompose the amplitude part and phase part of the wavefunction ansatz (Fig.~\ref{fig:summary}b)
 \begin{equation}
\Psi(x,\theta)=\pi(x)e^{i\phi(x)},
 \end{equation}
 where the conditional probabilities $\pi(x)=\prod_{k=1}^{N}|\pi_k(x_k|x_{k-1},...,x_1)|$ define a joint probability function $\pi : \{ 0,1\}^{N}\rightarrow[0,\infty)$ for input configuration strings $x$. The phase $\phi:\{ 0,1\}^N\rightarrow[0,2\pi)$ of the wavefunction directly by feeding the input $x$ into a two-layer multilayer perceptron (MLP). $\theta\in \mathbb{R}^{d}$ denotes the concatenation of parameters describing the neural networks $\pi$ and $\phi$. Our ansatz is autoregressive since only the amplitude contributes to the probability $|\pi(x)|^2$
and the amplitude part is autoregressive
by design, therefore we can use very efficient sampling algorithms designed for autoregressive models.

The autoregressive sampling generates one sample per run, performing $N$ local samplings in the process. To improve efficiency, Ref.~\cite{barrett2022autoregressive} introduces the batch autoregressive sampling (BAS) algorithm, which generates a batch of samples at each local sampling step instead of a single sample. Specifically, in the first step, $N_{\text{s}}$ samples (potentially as large as $10^{12}$) are generated from the probability distribution $\pi(x)$, retaining only unique samples with their weights. At the $k$-th step, the conditional probability $\pi_k(x_k|x_{k-1},...,x_1)$ is computed for each unique sample from the previous step (weighted by $w_{x_{k-1},...,x_1}$), and local samplings are performed to generate exactly $w_{x_{k-1},...,x_1}$ samples. Intermediate samples with zero weights are pruned at each layer, ultimately producing $N_{\text{s}}$ samples at the $N$-th step, stored as $N_{\text{u}}$ unique samples with weights. The BAS algorithm, illustrated in Fig.~\ref{fig:summary}a (middle), has a computational cost of $\sum_{n=1}^{N} k^2 N_{\text{u},k}$, where $N_{\text{u},k}$ is the number of unique samples at the $k$-th step. Assuming $N_{\text{u},k}\approx N_{\text{u}}$ (typically $N_{\text{u}}<10^{6}$ for molecular systems with $N>100$), the total cost of BAS scales as $O(N_{\text{u}}N^3/3)$, independent of $N_{\text{s}}$.

\textbf{Parallel sampling.} 
The first challenge for NQS lies in the sampling process. Autoregressive neural networks enable exact sampling without pre-thermalization or discarding intermediate samples to avoid auto-correlations, making them more efficient than traditional Monte Carlo methods. The BAS process can be visualized as a tree, with each layer representing a local sampling step~\cite{wu2023nnqs}, as shown in Fig.~\ref{fig:summary}a (middle). To parallelize BAS, we first perform serial sampling across all processes using the same random seed for the first $k$ steps, ensuring identical samples on each process. At the $k$-th step, unique samples are divided among $N_{\text{p}}$ processes, balancing the number of samples per process. The value of $k$ is dynamically determined by setting a threshold 
$N_{\text{u}}^*$ and choosing the first step where the number of unique samples ($N_{\text{u},k}$) exceeds $N_\text{u}^*$. This parallel sampling scheme allows for the computation of molecules with more electrons compared to other NQS methods (see Table~\ref{tab:eg}).

At each step, leaves with zero occurrence are pruned. Additionally, a number-conserving constraint reduces the sample space by preserving the total spin-up ($N_{\uparrow}$) and spin-down ($N_{\downarrow}$) electron counts separately. During the Jordan-Wigner transformation, the $i$-th spatial orbital maps to two qubits at positions $2i-1$ and $2i$, representing spin-up and spin-down orbitals. For each sampled pair $x_i=s_{2i}s_{2i-1} \; (s_i\in \{ 0,1 \})$ , the constraint is imposed by regularizing the local conditional probabilities $\pi_i(x_i|x_{i-1},...,x_1)$ as~\cite{zhao2023scalable,wu2023nnqs}:
\begin{equation}
    \tilde{\pi}_i(x_i|x_{i-1},...,x_1)=
    \left\{
    \begin{array}{l}
    0,\text{if} \; \sum_{j=1}^{i} s_{2j-1} > N_{\uparrow} \\
    0, \text{if} \; \sum_{j=1}^{i} s_{2j} > N_{\downarrow}  \\
    \pi_i(x_i|x_{i-1},...,x_1),o.w.
    \end{array}
    \right. .
\end{equation}
The adjusted probabilities are normalized as $\tilde{\pi}_i(x_i|x_{i-1},...,x_1)/\sum_{i} \tilde{\pi}_i(x_i|x_{i-1},...,x_1)$. 

\begin{figure}[htbp]
    \centering
    \includegraphics[width=0.55\linewidth]{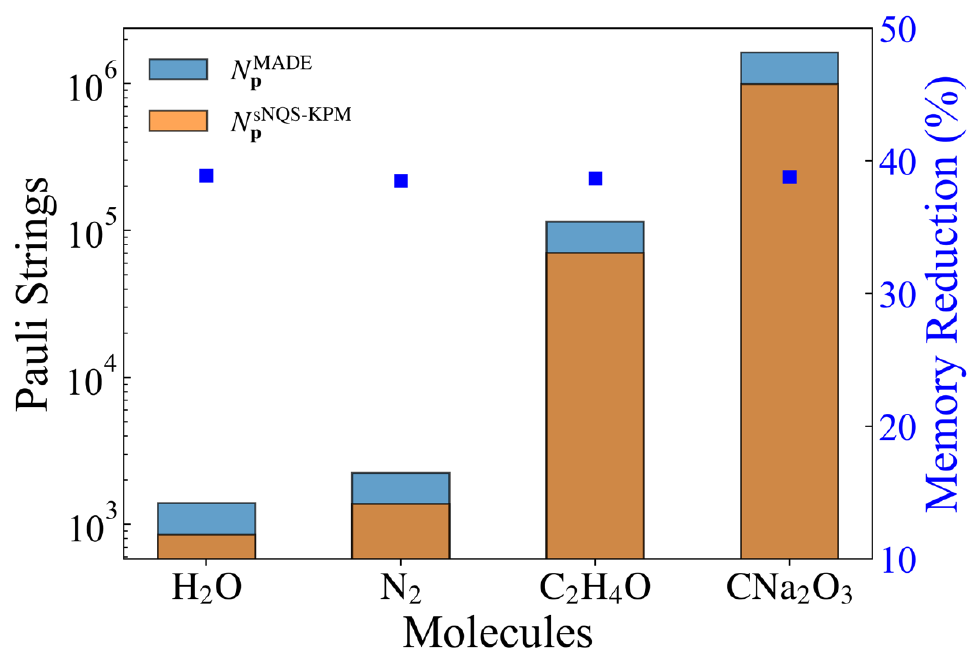}
    \caption{\textbf{Memory efficiency of sNQS-KPM.} Comparison of memory requirements for storing the Hamiltonian between sNQS-KPM and MADE from Ref.~\cite{zhao2023scalable} across various representative molecular systems in the STO-3G basis set.}
    \label{fig:mem}
\end{figure}

\textbf{GPU-supported local energy calculation.} The second challenge in NQS for electronic structure calculations is evaluating the local energy in Eq.~\ref{eqn:loss}, which is computationally expensive for molecular systems due to the $O(N^4)$ scaling of Pauli strings and memory issues from $O(N^4)$ possible $\mathbf{x}'$ coupled to an input $\mathbf{x}$. 
To address this, we implemented an efficient tensor representation of the second quantized spin Hamiltonian derived from chemical data, leveraging the sparsity of the matrix corresponding to a Pauli operator product $\hat{P}_{\mathbf{p}}$. Specifically, for each row index $x\in \{ 0,1 \}^{N}$, there is exactly one column index $x^{\mathbf{p}}\in \{ 0,1 \}^N$ with a nonzero matrix entry $\expval{x|\hat{P}_{\mathbf{p}}|x^{\mathbf{p}}}$. Denoting the subset of Pauli strings with nonzero coefficients as $S=\{ \mathbf{p} \in \{ I,x,y,z \} :h_{\mathbf{p}}\neq0 \}$, the local energy simplifies to:
\begin{equation}
    l_\theta (x) = \sum_{\mathbf{p}\in S} h_{\mathbf{p}} \expval{x|\hat{P}_{\mathbf{p}} |x^{\mathbf{p}}} \frac{\Psi(x^{\mathbf{p}},\theta)}{\Psi(x,\theta)}.
    \label{eqn:rloss2}
\end{equation}
To efficiently compute this large summation $K=|S|$ , we developed a GPU-compatible approach outlined in Algorithm~\ref{alg:parallel-local-energy}. Key information from the molecular Hamiltonian, such as Pauli string indices and coefficients, is stored as tensors supporting GPU computation, as shown in Fig.~\ref{fig:summary}a. A string parser constructs an incidence matrix with coefficients, tracking indices of Pauli operators ($\sigma_x$, $\sigma_y$, $\sigma_z$) as tensors for efficient computation of $x^{\mathbf{p}}$ and the matrix element $\expval{x\big|\hat{P}_{\mathbf{p}}\big|x^{\mathbf{p}}}$.

\begin{algorithm}
\caption{Parallel tensor computation of local energy (batch size = $B$).}
\label{alg:parallel-local-energy}
\begin{algorithmic}[1]
\State \textbf{Input:} Batch of bit strings $\big(\{x_b\} \in \{0,1 \}^{B \times N} \big)$, coefficient vector $\big(h_{\mathbf{p}}\big)$, unique incidence matrix $\big(M_{\text{u}}^{\{x,y \}}\in \{ 0,1 \}^{K\times N}\big)$, incidence matrix $\big(M^{\{y,z \}}\in \{ 0,1 \}^{K\times N}\big)$

\State \textbf{Output:} Batch of local energies $\{E_b\} \in \mathbb{C}^B$

\State \textbf{Parallel for} $b = 1$ to $B$ \textbf{in parallel}
    \State \hphantom{0000000} Compute ${x_b^{\mathbf{p}}}\in\{0,1 \}^{K\times N}$ as ${x_b^{\mathbf{p}}}=X_b\oplus M_{\text{u}}^{\{ x,y\}}$, where $X_b \in \{0,1\}^{K\times N}$ ($K$-fold replication of $x_b$)
    \State \hphantom{0000000} Compute amplitudes $\Psi(x_b)\in \mathbb{C}$ and $\{ \Psi(x_b^{\mathbf{p}}) \} \in \mathbb{C}^{K}$
    \State \hphantom{0000000} Compute $\bigg\{ \expval{x_b|\hat{P}_\mathbf{p}|x_b^{\mathbf{p}}} \bigg\} \in \mathbb{C}^{K} $ using Eq.~\ref{eqn:expval}
    \State \hphantom{0000000} Evaluate the sum in Eq.~\ref{eqn:rloss2} to get $E_b$
\State \textbf{End parallel for}

\end{algorithmic}
\end{algorithm}

In practice, $x^{\mathbf{p}}$ is computed by flipping the bits of $x$ based on the positions of $\sigma_x$ and $\sigma_y$ in $P_{\mathbf{p}}$. To enhance efficiency, the indices of $\sigma_x$, $\sigma_y$ operators ($M^{\{x,y\}}$) are pre-collected, storing only unique flips ($M_{\text{u}}^{\{x,y\}}$) and reorganizing related indices ($M^{\{y,z\}}$) and their occurrence counts (r)~\cite{wu2023nnqs}. The reorganization step is pre-computed and incurs negligible cost. This scheme reduces memory usage by approximately $40\%$ compared to MADE (see Fig.~\ref{fig:mem}), effectively shrinking the input size for the forward pass by eliminating redundant samples, which is especially beneficial for large-scale problems. The matrix element $\expval{x\big|\hat{P}_{\mathbf{p}}\big|x^{\mathbf{p}}}$ is given by: 
\begin{equation}
    \expval{x\big|\hat{P}_{\mathbf{p}}\big|x^{\mathbf{p}}} = \expval{x\Bigg|\prod_{i=1}^{N}\sigma_{p_i}\Bigg|x^{\mathbf{p}}} = (-i)^r \prod_{k:p_k \in \{y,z \}} (-1)^{x_k}
    \label{eqn:expval}
\end{equation}
where $r$ is the count of $\sigma_y$ occurrences ($r=\sum_{k=1}^n \mathbf{1}_{\{ p_k=y \}}$). Similar to before, the indices of $\sigma_y$, $\sigma_z$ operators are pre-collected, and the product is efficiently computed via the Hadamard product of the indices with $x$, followed by a reduction of entries. All computations are parallelized for GPU execution~\cite{barrett2022autoregressive,zhao2023scalable,wu2023nnqs}. 

\textbf{Absorption spectra of different molecules.} 
To validate the precision of our sNQS-KPM approach, we first compute the ground-state energies of several small-scale molecular systems and compare them with existing NQS results, as summarized in Table~\ref{tab:eg}. NAQS can handle molecules with up to 18 electrons, while MADE extends this limit to 36 electrons. In comparison, sNQS-KPM significantly pushes the boundary, enabling calculations for molecules with up to 52 electrons. The mean absolute errors (MAE) for each method are reported. Our method achieves comparable accuracy to NAQS (and surpasses it for \(\text{C}_2\), \(\text{N}_2\), and \(\text{O}_2\)), while consistently outperforming MADE. This level of precision ensures FCI-level accuracy for the absorption spectra calculations, even in cases where direct FCI computation is infeasible. 

\begin{figure}[htbp]
    \centering
    \includegraphics[width=0.85\linewidth]{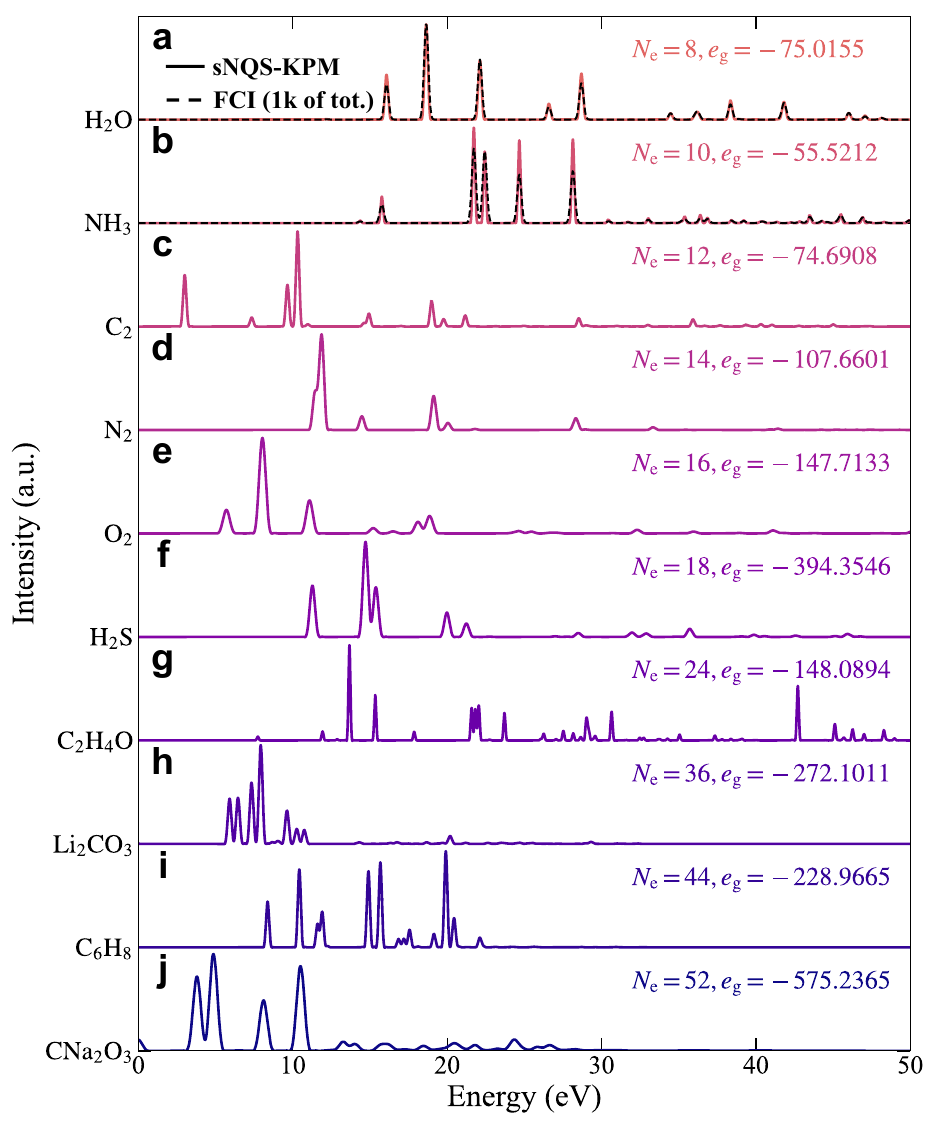}
    \caption{\textbf{Absorption spectra of various molecules computed using sNQS-KPM.} Molecules are arranged in order of increasing electron count. The black dotted line represents the absorption spectra of $\text{H}_2\text{O}$ and $\text{NH}_3$, obtained by diagonalizing the smallest 1000 eigenvalues and eigenvectors of the FCI Hamiltonian matrix (dimension \(2^n\), where \(n\) is the number of orbitals) on the STO-3G basis using the Lanczos algorithm.}
    \label{fig:full2}
\end{figure}

We show the performance of sNQS-KPM in calculating absorption spectra for various molecules sorted by electron count, as shown in Fig.~\ref{fig:full2}. For \(\text{H}_2\text{O}\) and \(\text{NH}_3\), FCI absorption spectra serve as baselines, derived by diagonalizing the smallest 1000 eigenvalues and eigenvectors of the FCI Hamiltonian matrix (out of \(2^n\), where \(n\) is the number of orbitals) due to computational limits. As illustrated in Figs.~\ref{fig:full2}a and \ref{fig:full2}b, sNQS-KPM accurately reproduces the peak positions of absorption spectra with high fidelity to FCI results. Minor discrepancies in relative peak heights arise from the limited number of eigenstates used in FCI, which decrease as more eigenstates are included.

Our method exhibits robust performance for molecules with 12 to 52 electrons, including \(\text{C}_2\), \(\text{N}_2\), \(\text{O}_2\), \(\text{H}_2\text{S}\), \(\text{C}_2\text{H}_4\text{O}\), \(\text{Li}_2\text{CO}_3\), \(\text{C}_6\text{H}_8\), and \(\text{CNa}_2\text{O}_3\). As molecule size increases, the rapidly growing number of terms in the electronic Hamiltonian poses challenges in computational cost and memory usage. Despite this, our algorithm efficiently scales to systems with up to 52 electrons, handling Hamiltonians with up to 1.6 million terms, while achieving state-of-the-art excited-state performance. Notably, in Fig.~\ref{fig:full2}j, a peak near relative energy of 0 indicates a degenerate ground state for \(\text{CNa}_2\text{O}_3\).

\begin{table}[h]
\caption{\textbf{Ground state energies (in Hartree) calculated by our method (sNQS-KPM).} The conventional HF and CCSD results, FCI results as well as existing NQS results including NAQS~\cite{barrett2022autoregressive} and MADE~\cite{zhao2023scalable} are shown for comparison. $N_{\text{e}}$ the total number of electrons (including spin up and spin down). The mean absolute errors (MAE) for various methods compared with FCI are also listed.}\label{tab:eg}%
\begin{tabular}{@{}cccccccc@{}}
\toprule
Molecule & $N_{\text{e}}$  & HF & CCSD & NAQS~\cite{barrett2022autoregressive} & MADE~\cite{zhao2023scalable} & Ours & FCI\\
\midrule
$\text{H}_2\text{O}$    & 8   & -74.9644  & -75.0154 & -75.0155 & -75.0155& -75.0155 & -75.0155\\
$\text{NH}_3$    & 10   & -55.4548 & -55.5209 & -55.5212 & -55.5212& -55.5212 & -55.5212\\
$\text{C}_2$    & 12   & -74.4209 & -74.6745 & -74.6889 & -74.6876 & -74.6908 & -74.6908\\
$\text{N}_2$    & 14   & -107.4990 & -107.6561 & -107.6595 & -107.6567& -107.6601 & -107.6602\\
$\text{O}2$    & 16   & -147.5513 & -147.7027 & -147.7127 & -147.7106& -147.7133 & -147.7133\\
$\text{H}_2\text{S}$    & 18   & -349.3114 & -394.3483 & -394.3546 & -394.3545& -394.3546 & -394.3546\\
$\text{C}_2\text{H}_4\text{O}$    & 24   & -148.0456 & -148.0798 & --- & -148.0827 & -148.0894& ---\\
$\text{Li}_2\text{CO}_3$    & 36   & -272.0976 & -272.0998 & --- & -272.1003& -272.1011 & ---\\
$\text{C}_6\text{H}_8$    & 44   & -228.9263 & -228.9586 & --- & ---& -228.9665 & ---\\
$\text{CNa}_2\text{O}_3$    & 52   & -575.0161 & -575.2295 & --- & ---& -575.2365 & ---\\
MAE(Hartree) & & & & $5.3 \times 10^{-4}$& $1.6 \times 10^{-3}$ & $1.7 \times 10^{-4}$& \\
\botrule
\end{tabular}
\end{table}

\textbf{Resolution of absorption spectra.} 
Our sNQS-KPM calculates the absorption spectrum using Chebyshev polynomial expansion. The number of polynomials required for the expansion is directly related to the resolution of the absorption spectrum. As discussed in Methods, to expand the $\delta$ function using Chebyshev polynomials, the eigenvalues of the Hamiltonian are typically rescaled to the range $[-1,1]$. This rescaling is necessary because Chebyshev polynomials within this range exhibit excellent properties such as orthogonality and recurrence relations, which enhance the stability and efficiency of the calculations. However, different molecules have different ranges of eigenvalues for their Hamiltonians, such as $[a,b]$. This necessitates different rescaling ranges. For instance, the eigenvalue range of the Hamiltonian of $\text{H}_2\text{O}$ is much smaller than that of $\text{CNa}_2\text{O}_3$. To achieve the same final absorption spectral resolution for both molecules, $\text{CNa}_2\text{O}_3$ requires a significantly larger number of moments ($N_\Omega$,  equivalent to the number of polynomials) compared to $\text{H}_2\text{O}$. In Figs.~\ref{fig:full2}a and \ref{fig:full2}j, $\text{H}_2\text{O}$ used $60,000$ moments, while $\text{CNa}_2\text{O}_3$ used $200,000$ moments. Despite this, the resolution of $\text{CNa}_2\text{O}_3$ appears slightly lower than that of $\text{H}_2\text{O}$.

\begin{figure}[htbp]
    \centering
    \includegraphics[width=0.96\linewidth]{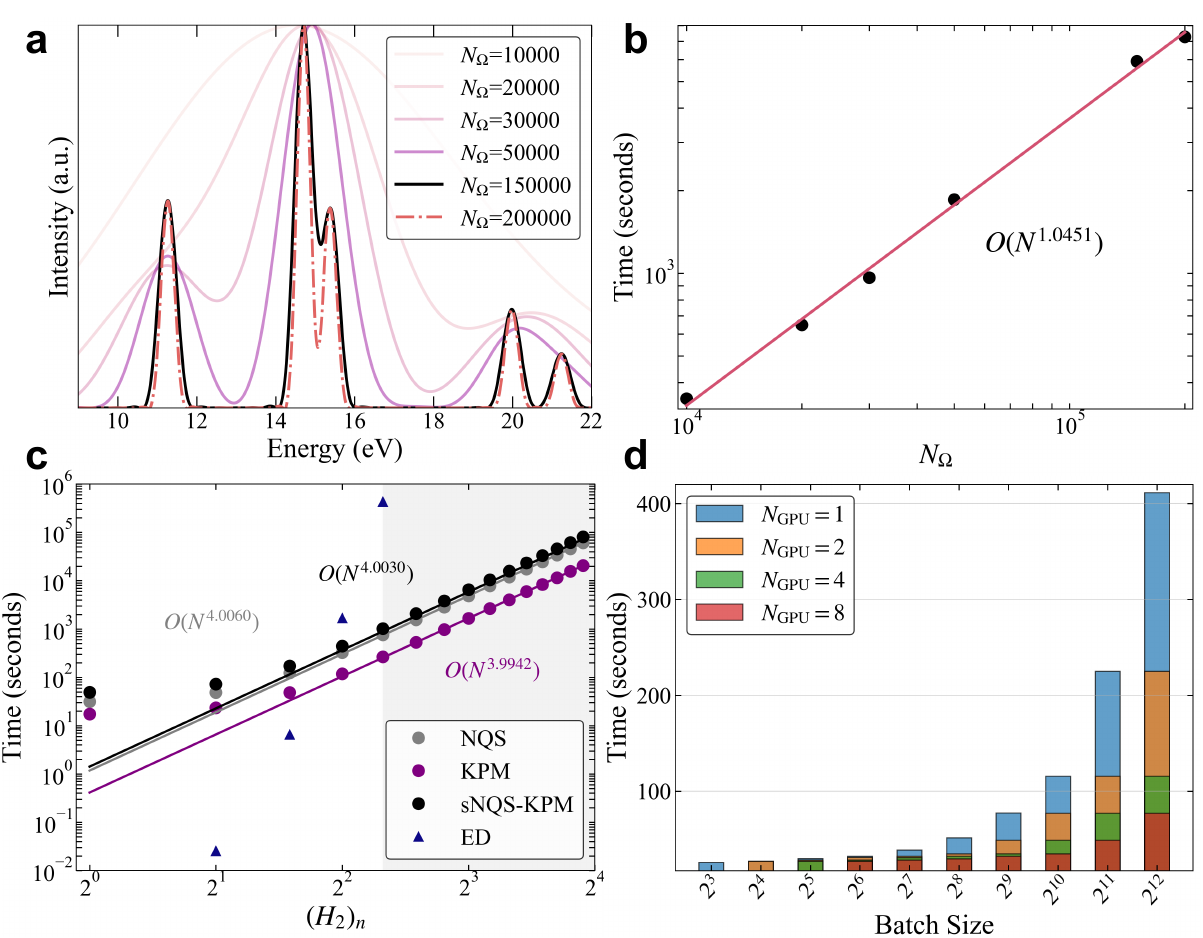}
    \caption{\textbf{Performance benchmark of sNQS-KPM.} \textbf{a,} Convergence of absorption spectral resolution for $\text{H}_2\text{S}$ as a function of the number of moments (\(N_\Omega\)). \textbf{b,} Relationship between computation time and the number of moments (\(N_\Omega\)) calculated. \textbf{c,} Comparison of computational time complexity for NQS, KPM, and the overall sNQS-KPM approach against exact diagonalization (ED) on extended $\text{H}_2$ chains. The benchmark is performed on a single V100 GPU, with NQS trained for $10^4$ steps and KPM calculating $10^4$ moments. \textbf{d,} Demonstration of near-optimal weak scaling for the $\text{O}_2$ molecule, running the algorithm for \(10^3\) iterations with varying batch sizes up to $2^{12}$ and reporting the elapsed time in seconds. Distributing batches across multiple GPUs significantly reduces runtime, while training with a fixed batch size per GPU maintains consistent runtimes across different configurations.}
    \label{fig:mom}
\end{figure}

We report the convergence of the absorption spectral resolution as a function of $N_\Omega$ in Fig.~\ref{fig:mom}a. We gradually increased $N_\Omega$ from $10,000$ to $200,000$, and the absorption spectrum of $\text{H}_2\text{S}$ evolved from a single peak to five distinct peaks within the energy range of 10 to 22 eV. To ensure convergence, we propose an empirical formula:
\begin{equation}
    N_\Omega=400|e_{\text{g}}|,
    \label{eqn:Nomega}
\end{equation}
where $e_{\text{g}}$ is the numerical value of the molecular ground state energy in atomic units, calculated using NQS. The computation time for the moments is approximately linearly related to the number of moments, as shown in Fig.~\ref{fig:full2}b. 

\textbf{Computational complexity.} 
We evaluated the time complexity of our method using \(\text{H}_2\) chains, as shown in Fig.~\ref{fig:full2}c. The tests were performed on a single V100 GPU, with 10,000 training steps using NQS and 10,000 moments calculated via KPM. By fitting the computation times for \(\text{H}_2\) chains containing 5 to 15 \(\text{H}_2\) molecules, we found that the NQS, KPM, and overall calculation times scale approximately $O(N^4)$ with the number of molecules in the chains. When the number of $\text{H}_2$ in the chain increases, the computational efficiency of sNQS-KPM is much higher than that of ED. Specifically, the time for ED to calculate the chain $(\text{H}_2)_5$ is already greater than that for sNQS-KPM to calculate the chain $(\text{H}_2)_{16}$.

To assess the effectiveness of our parallelization scheme, we examined the running time for the \(\text{O}_2\) molecule using \(10^3\) iterations across varying batch sizes, as illustrated in Fig.~\ref{fig:full2}d. The results show that the running time scales inversely with the number of GPUs, with doubling the GPUs reducing the time by roughly half. Furthermore, experiments with memory-saturated GPUs confirmed that the running time remains consistent across different GPU counts, demonstrating near-optimal weak scaling of our approach.

\section{Discussion}
This work presents a scalable neural quantum state-based kernel polynomial method (sNQS-KPM) for ab-initio absorption spectra, capable of achieving FCI-level accuracy even for systems with up to 52 electrons, where traditional FCI calculations become computationally infeasible. By integrating neural quantum states (NQS) with the kernel polynomial method (KPM), this approach leverages Chebyshev polynomial expansions to compute absorption spectra with high precision, while mitigating truncation errors through kernel smoothing techniques.  

Key innovations of this work include embedding physical priors into the network architecture to enhance the accuracy of quantum system modeling and implementing an efficient sampling strategy that scales with the number of unique configurations rather than the total batch size, thereby reducing computational overhead. Additionally, parallel computation of local energy across multiple GPUs is employed, distributing term-wise calculations without inter-term interactions and leveraging embarrassingly parallel algorithms to improve scalability for Hamiltonians with many terms. Furthermore, the moments in KPM are efficiently computed using recursive relationships, eliminating the need to store large numbers of eigenvectors for absorption spectra calculations, resulting in significant storage savings. Collectively, these advancements greatly extend the utility of ML-based methods for excited-state calculations in second-quantized quantum chemistry~\cite{choo2020fermionic,barrett2022autoregressive,zhao2023scalable,wu2023nnqs}.

Despite these advances, second-quantized methods are inherently constrained by the choice of basis set, posing challenges in scaling to larger basis sets akin to scaling to larger molecules. In contrast, first-quantized approaches, such as FermiNet~\cite{pfau2020ab} and PauliNet~\cite{hermann2020deep}, employ flexible basis sets and fewer determinants, achieving unconstrained accuracies at the cost of increased computational complexity. These methods rely on Markov chain Monte Carlo sampling, but their compatibility with second-quantized frameworks remains an open avenue for future investigation.

\section{Methods}
\textbf{Mapping fermionic Hamiltonians to the qubits.} 
We consider the second-quantized molecular electronic Hamiltonian~\cite{mcardle2020quantum}:  
\begin{equation}
    \hat{H} = \sum_{pq} h_{pq} \hat{a}_p^{\dagger} \hat{a}_q + \frac{1}{2} \sum_{pqrs} h_{pqrs} \hat{a}_p^{\dagger} \hat{a}_q^{\dagger} \hat{a}_r \hat{a}_s,
\end{equation}  
where \(\hat{a}_i^{\dagger}\) (\(\hat{a}_i\)) are fermionic creation (annihilation) operators for the \(i\)-th orbital, and \(h_{pq}\), \(h_{pqrs}\) are the one- and two-body integrals, respectively. Using the Jordan-Wigner encoding~\cite{jordan1993paulische}, these operators are mapped to qubit operators:  
\begin{equation}
    \hat{a}_j^{\dagger} \rightarrow \bigg( \prod_{i=0}^{j-1} \sigma_i^z \bigg) \sigma_j^{+}, \quad 
    \hat{a}_j \rightarrow \bigg( \prod_{i=0}^{j-1} \sigma_i^z \bigg) \sigma_j^-,
\end{equation}  
where the occupation state of the \(j\)-th fermionic mode is stored locally in the \(j\)-th qubit using \(\sigma_j^{\pm} = \sigma_j^x \pm \mathrm{i} \sigma_j^y\), while the parity is stored non-locally. Alternative mappings, such as parity~\cite{seeley2012bravyi} or Bravyi–Kitaev~\cite{bravyi2002fermionic} encodings, would require specific adjustments to encode physical priors but do not change the general form of the qubit Hamiltonian:  
\begin{equation}
    \hat{H}_{\text{Q}} = \sum_{\mathbf{p} \in \{I, x, y, z\}^N} h_{\mathbf{p}} \hat{P}_{\mathbf{p}},
    \label{eqn:HQ}
\end{equation}  
where \(\hat{P}_{\mathbf{p}} := \prod_{i=1}^N \sigma_{p_i}\) is a tensor product of Pauli matrices, and the coefficients \(h_{\mathbf{p}} = \frac{1}{2^N} \text{Tr}(\hat{H}_{\text{Q}} \hat{P}_{\mathbf{p}})\) are real-valued.

\textbf{Variational Monte Carlo optimization.} 
Following the standard neural-network variational Monte Carlo approach~\cite{choo2020fermionic}, we define trial wavefunctions whose complex amplitudes relative to the standard basis are computed by a neural network, parametrized by \(\theta \in \mathbb{R}^d\). Given a function  
\begin{equation}
    \Psi: \{0,1\}^N \times \mathbb{R}^d \to \mathbb{C},
\end{equation}
which is differentiable in \(\theta\), the associated family of trial quantum states is expressed as:  
\begin{equation}
    \ket{\Psi_\theta} = \sum_{x \in \{0,1\}^N} \Psi(x, \theta) \ket{x},
\end{equation}
where \(\ket{x} := \prod_{i=1}^N \ket{x_i}\) represents the standard basis vectors. The wavefunction \(\Psi(x, \theta)\) is normalized for all \(\theta\), and sampling from the probability distribution \(\pi_\theta := |\Psi(x, \theta)|^2\) is computationally feasible.  

Using \(\ket{\Psi_\theta}\) as a trial vector in the Rayleigh quotient for \(\hat{H}_{\text{Q}}\), we obtain a differentiable objective function:  
\begin{equation}
    \mathcal{E}(\theta) = \frac{1}{2} \expval{\Psi_\theta | \hat{H}_{\text{Q}} | \Psi_\theta},
\end{equation}
which upper bounds the minimal eigenvalue \(\lambda_{\text{min}}(\hat{H}_{\text{Q}})\) per the Rayleigh–Ritz principle. The objective can be estimated via Monte Carlo sampling:  
\begin{equation}
    \mathcal{E}(\theta) = \frac{1}{2} \mathbb{E}[l_\theta(x)], \quad l_\theta(x) = \frac{\expval{x | \hat{H}_{\text{Q}} | \Psi_\theta}}{\Psi(x, \theta)}, \; x \sim \pi_\theta,
    \label{eqn:loss}
\end{equation}
where \(l_\theta(x)\), the local energy, is crucial for both updating \(\theta\) and estimating \(\mathcal{E}(\theta)\). Stochastic gradient optimization minimizes \(\mathcal{E}\), with gradients estimated as:  
\begin{equation}
    \nabla \mathcal{E}(\theta) = \text{Re} \mathbb{E}[(l_\theta(x) \mathbf{1} - \mathbf{b}) \bar{\sigma}_\theta(x)], \quad \sigma_\theta(x) = \frac{\nabla_\theta \Psi(x, \theta)}{\Psi(x, \theta)}, \; x \sim \pi_\theta,
\end{equation}
where \(\mathbf{b}\) is a baseline matrix, often chosen as \(\mathbf{b} = \mathbb{E}[l_\theta(x)]\).  

Computing \(x \to l_\theta(x)\) for a minibatch of size \(B\) has complexity \(O(BK)\), where \(K\) is the number of terms in the Hamiltonian expansion (Eq.~\ref{eqn:HQ}). Although \(K \ll 4^N\), the sparsity does not prevent out-of-memory (OOM) issues for large molecules, as \(4^N\) grows exponentially with \(N\). For instance, Sodium Carbonate (\(N = 76\)) has \(4^N \approx 5.7 \times 10^{45}\) terms, while \(K = 1,625,991\). With a modest batch size \(B = 1024\), computing local energy involves a forward pass of 1.7 billion samples with input size 76, creating a significant computational bottleneck. Large batch sizes are essential for accurate Monte Carlo approximations of \(\mathcal{E}(\theta)\), directly impacting algorithm performance. Thus, efficient sampling and local energy computation are critical for scaling neural-network variational methods to large molecules.

\textbf{Parallel algorithm implementation.}
In quantum chemistry applications, even small molecules can have Hamiltonians with thousands of terms, causing severe OOM issues on existing VMC platforms. To address this, our pipeline tensorizes the molecular Hamiltonian to optimize memory usage. Local energy is computed term-wise with no inter-term interaction, enabling embarrassingly parallel algorithms for Hamiltonians with many terms. We tackle this bottleneck by applying a sampling parallelization strategy where identical model copies on multiple GPUs generate a few samples per unit. These independent samples are then combined to construct an accurate expectation estimate. Additionally, duplicate configurations are removed locally on each GPU before the forward pass to further reduce memory usage.  

The energy expectation is approximated as:  
\begin{equation}
    \frac{\expval{\Psi_\theta |\hat{H}_\text{Q}|\Psi_\theta}}{\expval{\Psi_\theta|\Psi_\theta}} \approx \frac{1}{B} \sum_{i=1}^{B} \sum_{\mathbf{p}\in S} h_{\mathbf{p}} \expval{x_i|P_{\mathbf{p}}|x_i^{\mathbf{p}}} \frac{\Psi(x_i^{\mathbf{p}},\theta)}{\Psi(x_i,\theta)}.
\end{equation}
We distribute the \(BK\) summands across \(L\) GPUs, each handling a partial sum of size \(BK/L\). Each GPU computes local gradients using forward and backward passes, and model parameters are updated by averaging these local gradients. To avoid OOM issues for larger molecules, we use gradient accumulation~\cite{lin2017deep}, splitting batches into smaller mini-batches before each update.

\textbf{KPM for absorption spectra.}
The KPM approximates physical quantities using polynomial expansions, with Chebyshev polynomials being the most commonly applied due to their efficiency. To mitigate truncation errors, modified kernels like the Jackson kernel~\cite{weisse2006kernel} are used, ensuring smooth approximations of functions such as the Dirac delta:  
\begin{equation}
    g_n = \frac{1}{N+1} \bigg[ (N-n-1)\cos\frac{\pi n}{N+1} + \sin\frac{\pi n}{N+1}\cot\frac{\pi}{N+1} \bigg].
    \label{eqn:jackson}
\end{equation}
For two operators \(A\) and \(B\), the dynamical correlation function is defined as:  
\begin{equation}
    \text{Im}[C_{AB}(\omega)] = \lim_{\epsilon \to 0} \bra{0} A \frac{1}{\omega + i\epsilon + H} B \ket{0},
\end{equation}
which, assuming real product \(\bra{0}A\ket{k}\bra{k}B\ket{0}\), simplifies to:  
\begin{equation}
    C_{AB}(\omega) = -\pi \sum_{k=0}^{D-1} \expval{0|A|k} \expval{k|B|0} \delta(\omega + E_k),
\end{equation}
where \(\ket{k}\) and \(E_k\) are the eigenstates and eigenvalues of \(H\), and \(\ket{0}\) is the ground state.  
After rescaling \(H \to \tilde{H}\) and \(\omega \to \tilde{\omega}\), \(\text{Im}[C_{AB}(\omega)]\) is expanded in Chebyshev polynomials:  
\begin{equation}
    \text{Im}[C_{AB}(\omega)] = -\frac{1}{\sqrt{1-\tilde{\omega}^2}} \bigg( \tilde{\Omega}_0 + 2 \sum_{n=1}^\infty \tilde{\Omega}_n T_n(\tilde{\omega}) \bigg),
\end{equation}
where \(\tilde{\Omega}_n = g_n \Omega_n\) and moments \(\Omega_n\) are computed as:  
\begin{equation}
    \Omega_n = \bra{0} A T_n(\tilde{H}) B \ket{0}.
\end{equation}

For dipole autocorrelation functions, with \(\hat{\mu} = \hat{\mu}^\dagger\), moments are:  
\begin{equation}
    \Omega_n = \bra{0} \mu^\dagger T_n(\tilde{H}) \mu \ket{0} = \bra{\alpha} T_n(\tilde{H}) \ket{\alpha}, 
\end{equation}
where \(\ket{\alpha} = \hat{\mu} \ket{0}\). The dipole operator \(\hat{\mu}\), in second-quantized form, is:  
\begin{equation}
    \hat{\mu} = -\sum_{pq} d_{pq} [\hat{c}_{p\uparrow}^\dagger \hat{c}_{q\uparrow} + \hat{c}_{p\downarrow}^\dagger \hat{c}_{q\downarrow}] - \hat{\mu}_\mathrm{c},
\end{equation}
with \(d_{pq}\) derived from the integral of the position operator \(\hat{\mathbf{r}}\) over molecular orbitals \(\phi\):  
\begin{equation}
    d_{pq} = \int dr \, \phi_p^*(r) \hat{\mathbf{r}} \phi_q(r),
\end{equation}
and core contributions \(\hat{\mu}_\mathrm{c}\):  
\begin{equation}
    \hat{\mu}_\mathrm{c} = 2 \sum_{i=1}^{N_\mathrm{core}} d_{ii}.
\end{equation}

Starting from \(\ket{\alpha}\), states \(\ket{\alpha_n} = T_n(\tilde{H}) \ket{\alpha}\) are iteratively constructed using the recursion:  
\begin{equation}
    \begin{aligned}
    \ket{\alpha_0} &= \ket{\alpha}, \\
    \ket{\alpha_1} &= \tilde{H} \ket{\alpha_0}, \\
    \ket{\alpha_{n+1}} &= 2\tilde{H} \ket{\alpha_n} - \ket{\alpha_{n-1}}.
\end{aligned}
\end{equation}

Moments are then computed as:  
\begin{equation}
    \begin{aligned}
    \Omega_{2n} &= 2\expval{\alpha_n | \alpha_n} - \expval{\alpha_0 | \alpha_0}, \\
    \Omega_{2n+1} &= 2\expval{\alpha_{n+1} | \alpha_n} - \expval{\alpha_1 | \alpha_0}.
\end{aligned}
\end{equation}
 
\textbf{Training details.}
Training follows the standard variational Monte Carlo approach described in the main text. Parameters are optimized using the Adam algorithm~\cite{kingma2014adam} with an initial learning rate of \(5 \times 10^{-3}\), reduced to \(5 \times 10^{-4}\) halfway through. The batch size, initially set to \(10^6\), is dynamically adjusted to maintain 10,000–100,000 unique, physically valid samples, with a maximum batch size of \(10^{12}\)~\cite{barrett2022autoregressive}.

\section{Data availability}
The exact molecular data generated, along with a notebook to reproduce these steps, can be found in the supporting code at \url{https://github.com/Weitheskmt/sNQS-KPM}.

\section{Code availability}
Source code to reproduce the reported results can be found at \url{https://github.com/Weitheskmt/sNQS-KPM}.

\section{Competing interests}
The authors declare no competing interests.

\section{Acknowledgments}
W.D. acknowledges the support from National Natural Science Foundation of China (No. 22361142829 and No. 22273075) and Zhejiang Provincial Natural Science Foundation (No. XHD24B0301). W.L. thanks Yu Wang
for useful discussions.

\bibliography{ref}


\begin{thebibliography}{45}
\ifx \bisbn   \undefined \def \bisbn  #1{ISBN #1}\fi
\ifx \binits  \undefined \def \binits#1{#1}\fi
\ifx \bauthor  \undefined \def \bauthor#1{#1}\fi
\ifx \batitle  \undefined \def \batitle#1{#1}\fi
\ifx \bjtitle  \undefined \def \bjtitle#1{#1}\fi
\ifx \bvolume  \undefined \def \bvolume#1{\textbf{#1}}\fi
\ifx \byear  \undefined \def \byear#1{#1}\fi
\ifx \bissue  \undefined \def \bissue#1{#1}\fi
\ifx \bfpage  \undefined \def \bfpage#1{#1}\fi
\ifx \blpage  \undefined \def \blpage #1{#1}\fi
\ifx \burl  \undefined \def \burl#1{\textsf{#1}}\fi
\ifx \doiurl  \undefined \def \doiurl#1{\url{https://doi.org/#1}}\fi
\ifx \betal  \undefined \def \betal{\textit{et al.}}\fi
\ifx \binstitute  \undefined \def \binstitute#1{#1}\fi
\ifx \binstitutionaled  \undefined \def \binstitutionaled#1{#1}\fi
\ifx \bctitle  \undefined \def \bctitle#1{#1}\fi
\ifx \beditor  \undefined \def \beditor#1{#1}\fi
\ifx \bpublisher  \undefined \def \bpublisher#1{#1}\fi
\ifx \bbtitle  \undefined \def \bbtitle#1{#1}\fi
\ifx \bedition  \undefined \def \bedition#1{#1}\fi
\ifx \bseriesno  \undefined \def \bseriesno#1{#1}\fi
\ifx \blocation  \undefined \def \blocation#1{#1}\fi
\ifx \bsertitle  \undefined \def \bsertitle#1{#1}\fi
\ifx \bsnm \undefined \def \bsnm#1{#1}\fi
\ifx \bsuffix \undefined \def \bsuffix#1{#1}\fi
\ifx \bparticle \undefined \def \bparticle#1{#1}\fi
\ifx \barticle \undefined \def \barticle#1{#1}\fi
\bibcommenthead
\ifx \bconfdate \undefined \def \bconfdate #1{#1}\fi
\ifx \botherref \undefined \def \botherref #1{#1}\fi
\ifx \url \undefined \def \url#1{\textsf{#1}}\fi
\ifx \bchapter \undefined \def \bchapter#1{#1}\fi
\ifx \bbook \undefined \def \bbook#1{#1}\fi
\ifx \bcomment \undefined \def \bcomment#1{#1}\fi
\ifx \oauthor \undefined \def \oauthor#1{#1}\fi
\ifx \citeauthoryear \undefined \def \citeauthoryear#1{#1}\fi
\ifx \endbibitem  \undefined \def \endbibitem {}\fi
\ifx \bconflocation  \undefined \def \bconflocation#1{#1}\fi
\ifx \arxivurl  \undefined \def \arxivurl#1{\textsf{#1}}\fi
\csname PreBibitemsHook\endcsname

\bibitem[\protect\citeauthoryear{Friesner}{2005}]{friesner2005ab}
\begin{barticle}
\bauthor{\bsnm{Friesner}, \binits{R.A.}}:
\batitle{Ab initio quantum chemistry: Methodology and applications}.
\bjtitle{Proceedings of the National Academy of Sciences}
\bvolume{102}(\bissue{19}),
\bfpage{6648}--\blpage{6653}
(\byear{2005})
\end{barticle}
\endbibitem

\bibitem[\protect\citeauthoryear{Hermann et~al.}{2023}]{hermann2023ab}
\begin{barticle}
\bauthor{\bsnm{Hermann}, \binits{J.}},
\bauthor{\bsnm{Spencer}, \binits{J.}},
\bauthor{\bsnm{Choo}, \binits{K.}},
\bauthor{\bsnm{Mezzacapo}, \binits{A.}},
\bauthor{\bsnm{Foulkes}, \binits{W.M.C.}},
\bauthor{\bsnm{Pfau}, \binits{D.}},
\bauthor{\bsnm{Carleo}, \binits{G.}},
\bauthor{\bsnm{No{\'e}}, \binits{F.}}:
\batitle{Ab initio quantum chemistry with neural-network wavefunctions}.
\bjtitle{Nature Reviews Chemistry}
\bvolume{7}(\bissue{10}),
\bfpage{692}--\blpage{709}
(\byear{2023})
\end{barticle}
\endbibitem

\bibitem[\protect\citeauthoryear{Bv and Hydraulics}{2000}]{bv2000absorption}
\begin{botherref}
\oauthor{\bsnm{Bv}, \binits{D.C.}},
\oauthor{\bsnm{Hydraulics}, \binits{D.}}:
Absorption spectroscopy.
Methods in Enzymologyvol
\textbf{5002011}
(2000)
\end{botherref}
\endbibitem

\bibitem[\protect\citeauthoryear{Zuehlsdorff and Isborn}{2019}]{zuehlsdorff2019modeling}
\begin{barticle}
\bauthor{\bsnm{Zuehlsdorff}, \binits{T.J.}},
\bauthor{\bsnm{Isborn}, \binits{C.M.}}:
\batitle{Modeling absorption spectra of molecules in solution}.
\bjtitle{International Journal of Quantum Chemistry}
\bvolume{119}(\bissue{1}),
\bfpage{25719}
(\byear{2019})
\end{barticle}
\endbibitem

\bibitem[\protect\citeauthoryear{Born and Oppenheimer}{1927}]{born1927quantentheorie}
\begin{barticle}
\bauthor{\bsnm{Born}, \binits{M.}},
\bauthor{\bsnm{Oppenheimer}, \binits{R.}}:
\batitle{Zur quantentheorie der molekeln.}
\bjtitle{Annalen der Physik}
\bvolume{389},
\bfpage{457}--\blpage{484}
(\byear{1927})
\end{barticle}
\endbibitem

\bibitem[\protect\citeauthoryear{Foulkes et~al.}{2001}]{foulkes2001quantum}
\begin{barticle}
\bauthor{\bsnm{Foulkes}, \binits{W.M.}},
\bauthor{\bsnm{Mitas}, \binits{L.}},
\bauthor{\bsnm{Needs}, \binits{R.}},
\bauthor{\bsnm{Rajagopal}, \binits{G.}}:
\batitle{Quantum monte carlo simulations of solids}.
\bjtitle{Reviews of Modern Physics}
\bvolume{73}(\bissue{1}),
\bfpage{33}
(\byear{2001})
\end{barticle}
\endbibitem

\bibitem[\protect\citeauthoryear{Szabo and Ostlund}{1996}]{szabo1996modern}
\begin{bbook}
\bauthor{\bsnm{Szabo}, \binits{A.}},
\bauthor{\bsnm{Ostlund}, \binits{N.S.}}:
\bbtitle{Modern Quantum Chemistry: Introduction to Advanced Electronic Structure Theory}.
\bpublisher{Courier Corporation},
\blocation{New York}
(\byear{1996})
\end{bbook}
\endbibitem

\bibitem[\protect\citeauthoryear{Knowles and Handy}{1984}]{knowles1984new}
\begin{barticle}
\bauthor{\bsnm{Knowles}, \binits{P.J.}},
\bauthor{\bsnm{Handy}, \binits{N.C.}}:
\batitle{A new determinant-based full configuration interaction method}.
\bjtitle{Chemical physics letters}
\bvolume{111}(\bissue{4-5}),
\bfpage{315}--\blpage{321}
(\byear{1984})
\end{barticle}
\endbibitem

\bibitem[\protect\citeauthoryear{Wei{\ss}e and Fehske}{2008}]{weisse2008exact}
\begin{botherref}
\oauthor{\bsnm{Wei{\ss}e}, \binits{A.}},
\oauthor{\bsnm{Fehske}, \binits{H.}}:
Exact diagonalization techniques.
Computational many-particle physics,
529--544
(2008)
\end{botherref}
\endbibitem

\bibitem[\protect\citeauthoryear{Szalay et~al.}{2012}]{szalay2012multiconfiguration}
\begin{barticle}
\bauthor{\bsnm{Szalay}, \binits{P.G.}},
\bauthor{\bsnm{Muller}, \binits{T.}},
\bauthor{\bsnm{Gidofalvi}, \binits{G.}},
\bauthor{\bsnm{Lischka}, \binits{H.}},
\bauthor{\bsnm{Shepard}, \binits{R.}}:
\batitle{Multiconfiguration self-consistent field and multireference configuration interaction methods and applications}.
\bjtitle{Chemical reviews}
\bvolume{112}(\bissue{1}),
\bfpage{108}--\blpage{181}
(\byear{2012})
\end{barticle}
\endbibitem

\bibitem[\protect\citeauthoryear{Marques and Gross}{2004}]{marques2004time}
\begin{barticle}
\bauthor{\bsnm{Marques}, \binits{M.A.}},
\bauthor{\bsnm{Gross}, \binits{E.K.}}:
\batitle{Time-dependent density functional theory}.
\bjtitle{Annu. Rev. Phys. Chem.}
\bvolume{55}(\bissue{1}),
\bfpage{427}--\blpage{455}
(\byear{2004})
\end{barticle}
\endbibitem

\bibitem[\protect\citeauthoryear{Kubo et~al.}{1991}]{kubo1991statistical}
\begin{botherref}
\oauthor{\bsnm{Kubo}, \binits{R.}},
\oauthor{\bsnm{Toda}, \binits{M.}},
\oauthor{\bsnm{Hashitsume}, \binits{N.}},
\oauthor{\bsnm{Kubo}, \binits{R.}},
\oauthor{\bsnm{Toda}, \binits{M.}},
\oauthor{\bsnm{Hashitsume}, \binits{N.}}:
Statistical mechanics of linear response.
Statistical Physics II: Nonequilibrium Statistical Mechanics,
146--202
(1991)
\end{botherref}
\endbibitem

\bibitem[\protect\citeauthoryear{Kubo}{1957}]{kubo1957statistical}
\begin{barticle}
\bauthor{\bsnm{Kubo}, \binits{R.}}:
\batitle{Statistical-mechanical theory of irreversible processes. i. general theory and simple applications to magnetic and conduction problems}.
\bjtitle{Journal of the physical society of Japan}
\bvolume{12}(\bissue{6}),
\bfpage{570}--\blpage{586}
(\byear{1957})
\end{barticle}
\endbibitem

\bibitem[\protect\citeauthoryear{Kubo et~al.}{1957}]{kubo1957statistical2}
\begin{barticle}
\bauthor{\bsnm{Kubo}, \binits{R.}},
\bauthor{\bsnm{Yokota}, \binits{M.}},
\bauthor{\bsnm{Nakajima}, \binits{S.}}:
\batitle{Statistical-mechanical theory of irreversible processes. ii. response to thermal disturbance}.
\bjtitle{Journal of the Physical Society of Japan}
\bvolume{12}(\bissue{11}),
\bfpage{1203}--\blpage{1211}
(\byear{1957})
\end{barticle}
\endbibitem

\bibitem[\protect\citeauthoryear{Kubo}{1966}]{kubo1966fluctuation}
\begin{barticle}
\bauthor{\bsnm{Kubo}, \binits{R.}}:
\batitle{The fluctuation-dissipation theorem}.
\bjtitle{Reports on progress in physics}
\bvolume{29}(\bissue{1}),
\bfpage{255}
(\byear{1966})
\end{barticle}
\endbibitem

\bibitem[\protect\citeauthoryear{Liu et~al.}{2023}]{liu2023predicting}
\begin{botherref}
\oauthor{\bsnm{Liu}, \binits{W.}},
\oauthor{\bsnm{Chen}, \binits{Z.-H.}},
\oauthor{\bsnm{Su}, \binits{Y.}},
\oauthor{\bsnm{Wang}, \binits{Y.}},
\oauthor{\bsnm{Dou}, \binits{W.}}:
Predicting rate kernels via dynamic mode decomposition.
The Journal of Chemical Physics
\textbf{159}(14)
(2023)
\end{botherref}
\endbibitem

\bibitem[\protect\citeauthoryear{Liu et~al.}{2024}]{liu2024memory}
\begin{botherref}
\oauthor{\bsnm{Liu}, \binits{W.}},
\oauthor{\bsnm{Su}, \binits{Y.}},
\oauthor{\bsnm{Wang}, \binits{Y.}},
\oauthor{\bsnm{Dou}, \binits{W.}}:
Memory kernel coupling theory: Obtain time correlation function from higher-order moments.
arXiv preprint arXiv:2407.01923
(2024)
\end{botherref}
\endbibitem

\bibitem[\protect\citeauthoryear{Iitaka}{1994}]{iitaka1994solving}
\begin{barticle}
\bauthor{\bsnm{Iitaka}, \binits{T.}}:
\batitle{Solving the time-dependent schr{\"o}dinger equation numerically}.
\bjtitle{Physical Review E}
\bvolume{49}(\bissue{5}),
\bfpage{4684}
(\byear{1994})
\end{barticle}
\endbibitem

\bibitem[\protect\citeauthoryear{Nys et~al.}{2024}]{nys2024ab}
\begin{barticle}
\bauthor{\bsnm{Nys}, \binits{J.}},
\bauthor{\bsnm{Pescia}, \binits{G.}},
\bauthor{\bsnm{Sinibaldi}, \binits{A.}},
\bauthor{\bsnm{Carleo}, \binits{G.}}:
\batitle{Ab-initio variational wave functions for the time-dependent many-electron schr{\"o}dinger equation}.
\bjtitle{Nature communications}
\bvolume{15}(\bissue{1}),
\bfpage{9404}
(\byear{2024})
\end{barticle}
\endbibitem

\bibitem[\protect\citeauthoryear{Wei{\ss}e et~al.}{2006}]{weisse2006kernel}
\begin{barticle}
\bauthor{\bsnm{Wei{\ss}e}, \binits{A.}},
\bauthor{\bsnm{Wellein}, \binits{G.}},
\bauthor{\bsnm{Alvermann}, \binits{A.}},
\bauthor{\bsnm{Fehske}, \binits{H.}}:
\batitle{The kernel polynomial method}.
\bjtitle{Reviews of modern physics}
\bvolume{78}(\bissue{1}),
\bfpage{275}--\blpage{306}
(\byear{2006})
\end{barticle}
\endbibitem

\bibitem[\protect\citeauthoryear{Vyazovskaya and Pupashenko}{2006}]{vyazovskaya2006normalizing}
\begin{botherref}
\oauthor{\bsnm{Vyazovskaya}, \binits{M.}},
\oauthor{\bsnm{Pupashenko}, \binits{N.}}:
On the normalizing multiplier of the generalized jackson kernel.
Mathematical Notes
\textbf{80}
(2006)
\end{botherref}
\endbibitem

\bibitem[\protect\citeauthoryear{Lykos and Pratt}{1963}]{lykos1963discussion}
\begin{barticle}
\bauthor{\bsnm{Lykos}, \binits{P.}},
\bauthor{\bsnm{Pratt}, \binits{G.}}:
\batitle{Discussion on the hartree-fock approximation}.
\bjtitle{Reviews of Modern Physics}
\bvolume{35}(\bissue{3}),
\bfpage{496}
(\byear{1963})
\end{barticle}
\endbibitem

\bibitem[\protect\citeauthoryear{Shavitt}{1977}]{shavitt1977method}
\begin{bchapter}
\bauthor{\bsnm{Shavitt}, \binits{I.}}:
\bctitle{The method of configuration interaction}.
In: \bbtitle{Methods of Electronic Structure Theory},
pp. \bfpage{189}--\blpage{275}.
\bpublisher{Springer},
\blocation{New York}
(\byear{1977})
\end{bchapter}
\endbibitem

\bibitem[\protect\citeauthoryear{Sherrill and Schaefer~III}{1999}]{sherrill1999configuration}
\begin{botherref}
\oauthor{\bsnm{Sherrill}, \binits{C.D.}},
\oauthor{\bsnm{Schaefer~III}, \binits{H.F.}}:
The configuration interaction method: Advances in highly correlated approaches
\textbf{34},
143--269
(1999)
\end{botherref}
\endbibitem

\bibitem[\protect\citeauthoryear{Bartlett and Musia{\l}}{2007}]{bartlett2007coupled}
\begin{barticle}
\bauthor{\bsnm{Bartlett}, \binits{R.J.}},
\bauthor{\bsnm{Musia{\l}}, \binits{M.}}:
\batitle{Coupled-cluster theory in quantum chemistry}.
\bjtitle{Reviews of Modern Physics}
\bvolume{79}(\bissue{1}),
\bfpage{291}--\blpage{352}
(\byear{2007})
\end{barticle}
\endbibitem

\bibitem[\protect\citeauthoryear{White}{1992}]{white1992density}
\begin{barticle}
\bauthor{\bsnm{White}, \binits{S.R.}}:
\batitle{Density matrix formulation for quantum renormalization groups}.
\bjtitle{Physical review letters}
\bvolume{69}(\bissue{19}),
\bfpage{2863}
(\byear{1992})
\end{barticle}
\endbibitem

\bibitem[\protect\citeauthoryear{White}{1993}]{white1993density}
\begin{barticle}
\bauthor{\bsnm{White}, \binits{S.R.}}:
\batitle{Density-matrix algorithms for quantum renormalization groups}.
\bjtitle{Physical review b}
\bvolume{48}(\bissue{14}),
\bfpage{10345}
(\byear{1993})
\end{barticle}
\endbibitem

\bibitem[\protect\citeauthoryear{Brabec et~al.}{2021}]{brabec2021massively}
\begin{barticle}
\bauthor{\bsnm{Brabec}, \binits{J.}},
\bauthor{\bsnm{Brandejs}, \binits{J.}},
\bauthor{\bsnm{Kowalski}, \binits{K.}},
\bauthor{\bsnm{Xantheas}, \binits{S.}},
\bauthor{\bsnm{Legeza}, \binits{{\"O}.}},
\bauthor{\bsnm{Veis}, \binits{L.}}:
\batitle{Massively parallel quantum chemical density matrix renormalization group method}.
\bjtitle{Journal of Computational Chemistry}
\bvolume{42}(\bissue{8}),
\bfpage{534}--\blpage{544}
(\byear{2021})
\end{barticle}
\endbibitem

\bibitem[\protect\citeauthoryear{Larsson et~al.}{2022}]{larsson2022chromium}
\begin{barticle}
\bauthor{\bsnm{Larsson}, \binits{H.R.}},
\bauthor{\bsnm{Zhai}, \binits{H.}},
\bauthor{\bsnm{Umrigar}, \binits{C.J.}},
\bauthor{\bsnm{Chan}, \binits{G.K.-L.}}:
\batitle{The chromium dimer: closing a chapter of quantum chemistry}.
\bjtitle{Journal of the American Chemical Society}
\bvolume{144}(\bissue{35}),
\bfpage{15932}--\blpage{15937}
(\byear{2022})
\end{barticle}
\endbibitem

\bibitem[\protect\citeauthoryear{Perez-Garcia et~al.}{2006}]{perez2006matrix}
\begin{botherref}
\oauthor{\bsnm{Perez-Garcia}, \binits{D.}},
\oauthor{\bsnm{Verstraete}, \binits{F.}},
\oauthor{\bsnm{Wolf}, \binits{M.M.}},
\oauthor{\bsnm{Cirac}, \binits{J.I.}}:
Matrix product state representations.
arXiv preprint quant-ph/0608197
(2006)
\end{botherref}
\endbibitem

\bibitem[\protect\citeauthoryear{Fischer and Igel}{2012}]{fischer2012introduction}
\begin{bchapter}
\bauthor{\bsnm{Fischer}, \binits{A.}},
\bauthor{\bsnm{Igel}, \binits{C.}}:
\bctitle{An introduction to restricted boltzmann machines}.
In: \bbtitle{Iberoamerican Congress on Pattern Recognition},
pp. \bfpage{14}--\blpage{36}
(\byear{2012}).
\bcomment{Springer}
\end{bchapter}
\endbibitem

\bibitem[\protect\citeauthoryear{Choo et~al.}{2020}]{choo2020fermionic}
\begin{barticle}
\bauthor{\bsnm{Choo}, \binits{K.}},
\bauthor{\bsnm{Mezzacapo}, \binits{A.}},
\bauthor{\bsnm{Carleo}, \binits{G.}}:
\batitle{Fermionic neural-network states for ab-initio electronic structure}.
\bjtitle{Nature communications}
\bvolume{11}(\bissue{1}),
\bfpage{2368}
(\byear{2020})
\end{barticle}
\endbibitem

\bibitem[\protect\citeauthoryear{Carleo and Troyer}{2017}]{carleo2017solving}
\begin{barticle}
\bauthor{\bsnm{Carleo}, \binits{G.}},
\bauthor{\bsnm{Troyer}, \binits{M.}}:
\batitle{Solving the quantum many-body problem with artificial neural networks}.
\bjtitle{Science}
\bvolume{355}(\bissue{6325}),
\bfpage{602}--\blpage{606}
(\byear{2017})
\end{barticle}
\endbibitem

\bibitem[\protect\citeauthoryear{Barrett et~al.}{2022}]{barrett2022autoregressive}
\begin{barticle}
\bauthor{\bsnm{Barrett}, \binits{T.D.}},
\bauthor{\bsnm{Malyshev}, \binits{A.}},
\bauthor{\bsnm{Lvovsky}, \binits{A.}}:
\batitle{Autoregressive neural-network wavefunctions for ab initio quantum chemistry}.
\bjtitle{Nature Machine Intelligence}
\bvolume{4}(\bissue{4}),
\bfpage{351}--\blpage{358}
(\byear{2022})
\end{barticle}
\endbibitem

\bibitem[\protect\citeauthoryear{Wu et~al.}{2023}]{wu2023nnqs}
\begin{bchapter}
\bauthor{\bsnm{Wu}, \binits{Y.}},
\bauthor{\bsnm{Guo}, \binits{C.}},
\bauthor{\bsnm{Fan}, \binits{Y.}},
\bauthor{\bsnm{Zhou}, \binits{P.}},
\bauthor{\bsnm{Shang}, \binits{H.}}:
\bctitle{Nnqs-transformer: an efficient and scalable neural network quantum states approach for ab initio quantum chemistry}.
In: \bbtitle{Proceedings of the International Conference for High Performance Computing, Networking, Storage and Analysis},
pp. \bfpage{1}--\blpage{13}
(\byear{2023})
\end{bchapter}
\endbibitem

\bibitem[\protect\citeauthoryear{Larochelle and Murray}{2011}]{larochelle2011neural}
\begin{bchapter}
\bauthor{\bsnm{Larochelle}, \binits{H.}},
\bauthor{\bsnm{Murray}, \binits{I.}}:
\bctitle{The neural autoregressive distribution estimator}.
In: \bbtitle{Proceedings of the Fourteenth International Conference on Artificial Intelligence and Statistics},
pp. \bfpage{29}--\blpage{37}
(\byear{2011}).
\bcomment{JMLR Workshop and Conference Proceedings}
\end{bchapter}
\endbibitem

\bibitem[\protect\citeauthoryear{Zhao et~al.}{2023}]{zhao2023scalable}
\begin{barticle}
\bauthor{\bsnm{Zhao}, \binits{T.}},
\bauthor{\bsnm{Stokes}, \binits{J.}},
\bauthor{\bsnm{Veerapaneni}, \binits{S.}}:
\batitle{Scalable neural quantum states architecture for quantum chemistry}.
\bjtitle{Machine Learning: Science and Technology}
\bvolume{4}(\bissue{2}),
\bfpage{025034}
(\byear{2023})
\end{barticle}
\endbibitem

\bibitem[\protect\citeauthoryear{Pfau et~al.}{2020}]{pfau2020ab}
\begin{barticle}
\bauthor{\bsnm{Pfau}, \binits{D.}},
\bauthor{\bsnm{Spencer}, \binits{J.S.}},
\bauthor{\bsnm{Matthews}, \binits{A.G.}},
\bauthor{\bsnm{Foulkes}, \binits{W.M.C.}}:
\batitle{Ab initio solution of the many-electron schr{\"o}dinger equation with deep neural networks}.
\bjtitle{Physical review research}
\bvolume{2}(\bissue{3}),
\bfpage{033429}
(\byear{2020})
\end{barticle}
\endbibitem

\bibitem[\protect\citeauthoryear{Hermann et~al.}{2020}]{hermann2020deep}
\begin{barticle}
\bauthor{\bsnm{Hermann}, \binits{J.}},
\bauthor{\bsnm{Sch{\"a}tzle}, \binits{Z.}},
\bauthor{\bsnm{No{\'e}}, \binits{F.}}:
\batitle{Deep-neural-network solution of the electronic schr{\"o}dinger equation}.
\bjtitle{Nature Chemistry}
\bvolume{12}(\bissue{10}),
\bfpage{891}--\blpage{897}
(\byear{2020})
\end{barticle}
\endbibitem

\bibitem[\protect\citeauthoryear{McArdle et~al.}{2020}]{mcardle2020quantum}
\begin{barticle}
\bauthor{\bsnm{McArdle}, \binits{S.}},
\bauthor{\bsnm{Endo}, \binits{S.}},
\bauthor{\bsnm{Aspuru-Guzik}, \binits{A.}},
\bauthor{\bsnm{Benjamin}, \binits{S.C.}},
\bauthor{\bsnm{Yuan}, \binits{X.}}:
\batitle{Quantum computational chemistry}.
\bjtitle{Reviews of Modern Physics}
\bvolume{92}(\bissue{1}),
\bfpage{015003}
(\byear{2020})
\end{barticle}
\endbibitem

\bibitem[\protect\citeauthoryear{Jordan and Wigner}{1993}]{jordan1993paulische}
\begin{bbook}
\bauthor{\bsnm{Jordan}, \binits{P.}},
\bauthor{\bsnm{Wigner}, \binits{E.P.}}:
\bbtitle{{\"U}ber Das Paulische {\"A}quivalenzverbot}.
\bpublisher{Springer},
\blocation{Verlag Berlin Heidelberg}
(\byear{1993})
\end{bbook}
\endbibitem

\bibitem[\protect\citeauthoryear{Seeley et~al.}{2012}]{seeley2012bravyi}
\begin{botherref}
\oauthor{\bsnm{Seeley}, \binits{J.T.}},
\oauthor{\bsnm{Richard}, \binits{M.J.}},
\oauthor{\bsnm{Love}, \binits{P.J.}}:
The bravyi-kitaev transformation for quantum computation of electronic structure.
The Journal of chemical physics
\textbf{137}(22)
(2012)
\end{botherref}
\endbibitem

\bibitem[\protect\citeauthoryear{Bravyi and Kitaev}{2002}]{bravyi2002fermionic}
\begin{barticle}
\bauthor{\bsnm{Bravyi}, \binits{S.B.}},
\bauthor{\bsnm{Kitaev}, \binits{A.Y.}}:
\batitle{Fermionic quantum computation}.
\bjtitle{Annals of Physics}
\bvolume{298}(\bissue{1}),
\bfpage{210}--\blpage{226}
(\byear{2002})
\end{barticle}
\endbibitem

\bibitem[\protect\citeauthoryear{Lin et~al.}{2017}]{lin2017deep}
\begin{botherref}
\oauthor{\bsnm{Lin}, \binits{Y.}},
\oauthor{\bsnm{Han}, \binits{S.}},
\oauthor{\bsnm{Mao}, \binits{H.}},
\oauthor{\bsnm{Wang}, \binits{Y.}},
\oauthor{\bsnm{Dally}, \binits{W.J.}}:
Deep gradient compression: Reducing the communication bandwidth for distributed training.
arXiv preprint arXiv:1712.01887
(2017)
\end{botherref}
\endbibitem

\bibitem[\protect\citeauthoryear{Kingma}{2014}]{kingma2014adam}
\begin{botherref}
\oauthor{\bsnm{Kingma}, \binits{D.P.}}:
Adam: A method for stochastic optimization.
arXiv preprint arXiv:1412.6980
(2014)
\end{botherref}
\endbibitem

\end{thebibliography}

\end{document}